# BERT and CNN integrated Neural Collaborative Filtering for Recommender Systems

Abdullah Al Munem
Department of Computer Science and Engineering
East West University
Dhaka, Bangladesh
abdullahalmunem@gmail.com

Sumona Yeasmin
Department of Computer Science and Engineering
East West University
Dhaka, Bangladesh
sumonasumu930@gmail.com

Mohammad Rezwanul Huq
Department of Computer Science and Engineering
East West University
Dhaka, Bangladesh
mrhuq@ewubd.edu

*Abstract*— Every day, a significant number of users visit the internet for different needs. The owners of a website generate profits from the user interaction with the contents or items of the website. A robust recommendation system can increase user interaction with a website by recommending items according to the user's unique preferences. BERT and CNN-integrated neural collaborative filtering (NCF) have been proposed for the recommendation system in this experiment. The proposed model takes inputs from the user and item profile and finds the user's interest. This model can handle numeric, categorical, and image data to extract the latent features from the inputs. The model is trained and validated on a small sample of the MovieLens dataset for 25 epochs. The same dataset has been used to train and validate a simple NCF and a BERT-based NCF model and compared with the proposed model. The proposed model outperformed those two baseline models. The obtained result for the proposed model is 0.72 recall and 0.486 Hit Ratio @ 10 for 799 users on the MovieLens dataset. This experiment concludes that considering both categorical and image data can improve the performance of a recommendation system.

## I. Introduction

The number of active users of the internet has been increasing. People generate a lot of data while browsing. To utilize the information, many online services, such as social media sites, streaming services, and e-commerce, have widely adopted recommender systems to increase user interaction and profit. Every user has some personal preferences, and recommending items based on user preferences can improve user experience and make them more engaged with the website. Effective recommendations can lead to significant profit increase [1,2].

The most commonly used approach for recommendation systems is Content-Based Filtering (CBF) and Collaborative Filtering (CF) [3]. Since CBF recommends items based on the item's features and the user's existing preferences, it cannot explore the users' new interests. CBF only considers item-item or user-user similarities for recommendations. To address the limitation of CBF, CF provides recommendations based on users' similarities between items and users simultaneously. Thus, CF can explore new interests for a particular user. CF can consider both explicit and implicit feedback from the user. But the data generated from CF has a high sparsity because most of the users will not interact with most of the items. Thus, it is computationally inefficient and resource hungry. To address this limitation, neural network-based CF called neural collaborative filtering (NCF) has been introduced which embeds the user-item pair and represents the latent features using dense vectors [4].

In this study, a hybrid NCF has been proposed to make the recommendation system more efficient and effective. The proposed model can consider a wide range of data that can be generated from the items and by the users. Bidirectional Encoder Representations from Transformers (BERT) and Convolutional Neural Network (CNN) has been interacted with the proposed model to handle both categorical and image data.

## II. Related Work

Traditional methods such as content-based recommendation systems [5] and Collaborative Filtering (CF) [6] have been used as a recommendation system for several years. Due to the extensive success of deep neural networks in various fields, it has been adopted in recommendation systems. Neural Collaborative Filtering (NCF) has shown tremendous improvement in recommending items based on user's unique preferences. Xiangnan et al. [4] proposed the NCF model, which outperforms the state-of-the-art models. Matrix Factorization (MF) is the most popular kind of collaborative filtering. Utilising a fixed inner product of the user-item matrix, user-item interactions are learned. Since calculating inner product is a computationally expensive task, NCF replaces the user-item inner product with a neural architecture. In order to learn about user-item interactions, NCF used a multi-layer perceptron and tried to articulate and generalize MF inside its framework. Since the proposed model of this study is developed based on the concept of NCF, this Related Works section only focused on NCF-based recommendation systems.

Juarto et al. [3] proposed BERT based NCF model for the news recommender system. They used the Microsoft news dataset to conduct this experiment. The Microsoft news dataset contains 5,475,542 user-news interactions among 50,000 users and 51,282 news. For embedding the news title and news content, they used Sentence BERT (SBERT) which is a modification of BERT and Label Encoder for embedding numerical data. They achieved a precision of 99.14%, f1-score of 95.69%, recall of 92.48%, ROC score of 98%, and hit ratio@10 of 74%.

Reinald et al. [7] proposed the BERT-Based Neural Collaborative Filtering and Fixed-Length Contiguous Tokens Explanation model to address the drawbacks of review-based recommender systems. They train this model on 4 different datasets and achieved average MSE of 0.8348, 0.8750, 0.9963, and 0.9764 on Toys and Games, Digital Music, Yelp-Dense, and Yelp-Sparse respectively. They used two input modules (User and Item) which pass through BERT (base) layers.

Another similar study was conducted by the authors of the paper [8]. They proposed a neural collaborative embedding model (NCEM) in their paper. They used item reviews to predict the user's rating. The inputs pass through the BERT encoder where the document embedding happens. They achieved MSE of 0.789, 0.816, 0.912, and 1.326 on Digital Music, Toys and Games, Yelp 2017, and Instant Video, respectively.

Boudiba et al. [9] proposed two Collaborative Filtering (CF) architectures, namely a CF-based Multilayer perceptron (MLP) Model and a CF-based Autoencoder Model. They used Word2vec and BERT and get the dense tag-based user and item representations. They performed this experiment on several MovieLens datasets. However, BERT based Neural CF-MLP Model outperformed most of the cases. Another paper [10] published by Boudiba et al. introduced one more model which is Graph-based NCF integrated with BERT for the recommendation system. This model is also trained on MovieLens datasets with the Experimental Settings mentioned in the paper [9]. The Graph-based NCF with BERT outperformed the other models.

There are some other significant works that have been done in this field. The authors of the papers [11], [12], [13], [14], and [15] used NCF for the recommendation systems. The authors of the paper [16] and [17] used Long short-term memory (LSTM) based NCF and the obtained result of both papers outperformed the NCF model. The advantage of using BERT over LSTM is BERT takes sequential inputs parallelly. Thus, BERT is faster than LSTM. BERT can also extract the contextual meanings of the words. To take advantage of BERT and CNN, a hybrid NCF integrated with BERT and CNN has been proposed in this paper.

## III. METHODOLOGY

A Hybrid Neural Collaborative Filtering (HNCF) architecture integrated with BERT and CNN has been used for the recommendation system. The proposed model can handle three types of data (numerical, categorical, and image) as inputs from the dataset. BERT and CNN architecture has been used to extract contextual information from categorical and image data respectively. A custom data generator has been used to preprocess the inputs and fit the inputs into the main memory. The dataset was split for training and validation purposes. Recall, and Hit Ratio @ 10 has been used to measure the model performance. Fig 1. shows the workflow of this experiment.

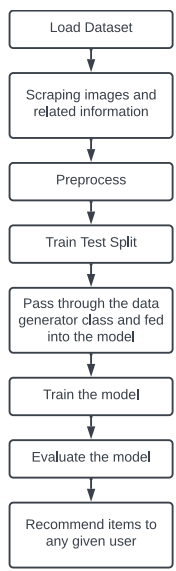

Fig. 1. The workflow of proposed methodology

### A. Dataset preparation

MovieLens 20M Dataset has been used to conduct this experiment. This da-taset has 6 CSV files (genome scores, genome tags, link, movie, rating, and tag) which contain information regarding movies and users. The CSV files are merged into a single CSV file based on the user id and movie id. The poster and related information of each movie has been scraped from IMDB and TMDB website using the IDs. IMDB and TMDB provide APIs for scraping the information. The path of the posters and the related information that is not present in the MovieLens 20M Dataset are inserted into the merged CSV file. By doing those steps, the dataset was prepared for preprocessing. The prepro-cessed dataset is available on Kaggle [18].

The preprocessed dataset contains 5 columns: User ID, Movie ID, Interac-tion, Images, and Processed Features. User ID and Movie ID contain the IDs of the users and movies respectively. Interaction represents whether the user has watched the corresponding movie or not. Interaction contains two unique val-ues: 0 (not watched) and 1 (watched). Interaction is the target column for the proposed model. Images columns contain the local path of the posters, and the Processed Features column contains movie information such as title, PG rat-ing, genres, spoken languages, movie description, tagline, keywords, casts, production companies, production countries, and tags name.

Random sampling has been used since the preprocessed dataset contains humongous records. It is impossible to process all those records with the lim-ited resources that have been used in this experiment. After the sampling, the dataset contains 1% records of the main dataset. The sampled dataset was fur-ther resampled by 0.02%. The sampled dataset was split into train and test da-tasets. The statistical summary of the sampled dataset is shown in table 1.

TABLE I.  STATISTICAL SUMMARY

| Parameters | Values |
|---|---|
| Train dataset | 15990 |
| Validation dataset | 3998 |
| Unique Users | 3033 |
| Unique Movies | 1762 |
| Class 1 in train dataset | 12826 |
| Class 0 in train dataset | 3164 |
| Class 1 in test dataset | 3145 |
| Class 0 in test dataset | 853 |

### B. Network Architecture

The proposed model takes four inputs: user id, movie id, movie features, and posters. The user id and movie id pass through the Keras Embedding layers. Keras Embedding layer extracts the latent features from the inputs and returns a sequence vector. After that, the sequence vector passes through a flattened layer. The movie features are fed into the BERT preprocess layer and the BERT encoder layer. BERT layers are used for extracting the contextual meaning from the contextual meaning. "BERT en uncased L-12 H-768 A-12" model has been used in this network where L is the number of stacked encoders, H is the hidden size and A is the number of heads in the MultiHead Attention layers. The BERT model returns a vector representation of the embedded features.

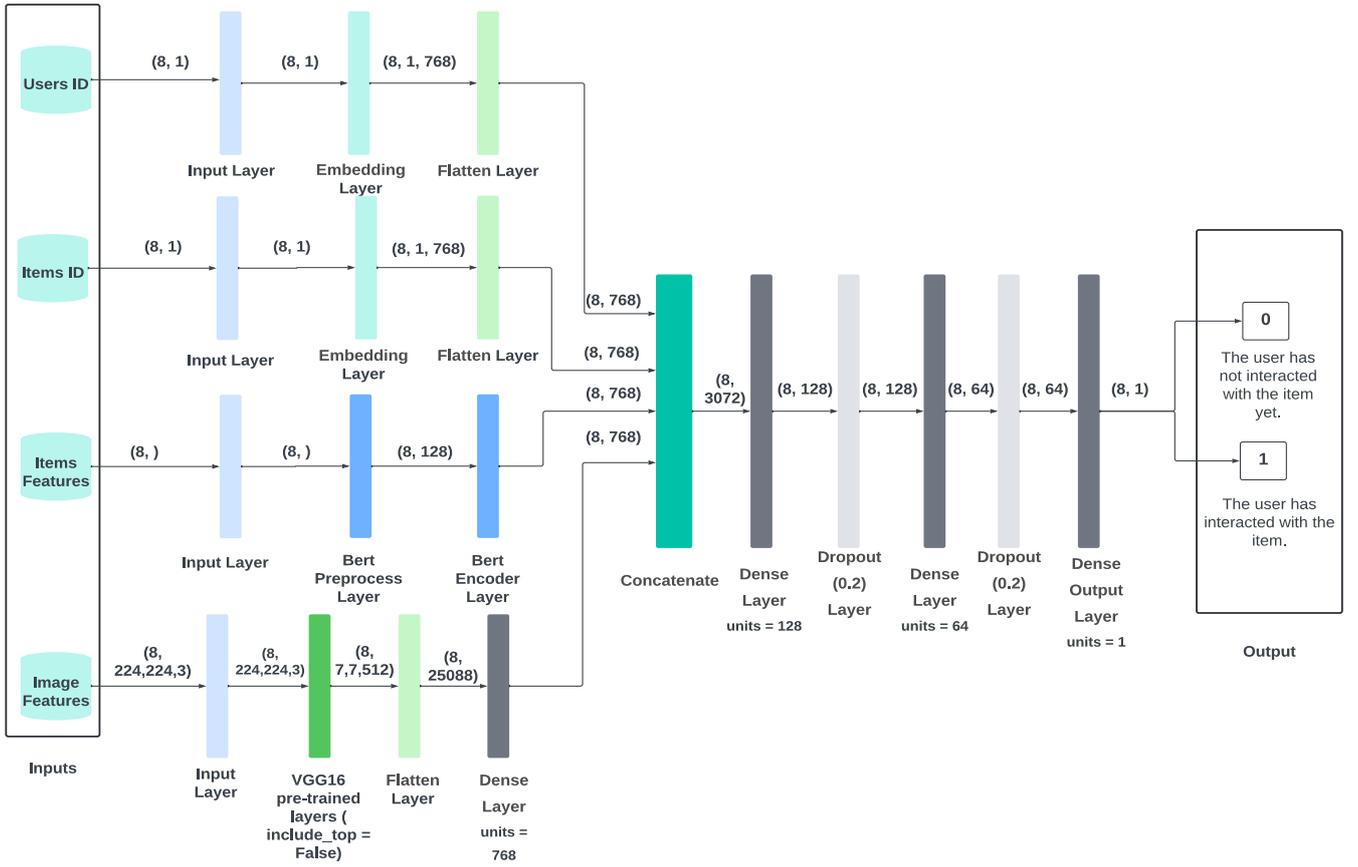

Fig. 2. BERT and CNN integrated Hybrid Neural Collaborative Filtering (HNCF) model architecture

The images are fed into VGG16 pre-trained CNN model to extract the contextual information from the posters. The top two layers of the VGG16 are discarded and froze the trainable parameters. The VGG16 model's CNN layers return a feature vector for the input image. The feature vectors pass through a flatten and dense layer.

After those steps, the user id embedded vector, movie id embedded vector, BERT embedded feature vector, and the VGG16 extracted image feature vector are concatenated and assigned into a single vector. The merged vector is fed as inputs into a multiple couples of dense and dropout layers. The final output dense layer gives the probability of user-item interaction. Fig 2. shows the network architecture of this proposed model.

### C. Data Pre-processing

The dataset contains three types of data: numerical (user id, movie id, interaction), categorical (movie features), and image (posters). The categorical data and images had been preprocessed before being fed into the model.

The steps for pre-processing the categorical data are following: (1) The text of the movie features is transformed into lowercase. (2) The special characters are removed and stop words are removed from the features. (3) Pass the pre-processed features into the BERT preprocess layers.

The steps for pre-processing the image are following: (1) Resize the image in-to a specific shape (224,224,3). (2) Convert the image into a 3D Numpy ar-ray. (3) Rescale/Normalize the image by dividing 255; since the highest in-tensity value of the image is 255.

### D. Experimental Setup

The model is developed using TensorFlow and the experiment is performed in Kaggle. Kaggle notebook provides 72GB HDD storage, 13GB RAM along with CPU, 14GB GPU for 40 hours per week, and 12 hours of the active session. It is not possible to train the model for more than 12 hours using Kaggle. Thus, this experiment took a sample of a faction of the main dataset. The values of the hyperparameters are shown in table 2.

TABLE II. THE SELECTED VALUES OF THE HYPERPARAMETERS

| Hyperparameters | Value(s) | |
|---|---|---|
| Class Mode | Numerical | |
| Transfer Learning Weights | ImageNet | |
| Validation Split | 20% | |
| Pooling | Max-Pooling | |
| Activation | Hidden Layers | ReLu |
| | Output Layer | Sigmoid |
| Optimizer | Adam | |
| Learning Rate | 0.001 | |
| Loss | Binary Cross Entropy | |
| Metrics | Hit Ratio @ 10, and Recall | |
| Epoch | 25 | |
| Batch size | 8 | |

## IV. Performance Evaluation

Depending on the data generated by user interactions with items, the recommendation system employs multiple evaluation methods. In this experiment, we used implicit data to demonstrate the user interactions with items. For example, if a user watches a movie, then it is symbolized as 1 otherwise 0. To evaluate the implicit data, we used Hit Ratio @ 10, and recall as evaluation metrics. The recall is the ratio between true positive and predicted results. It represents how accurately the model can predict whether a user has interacted with the item or not. False Negative has a significant value for the recommendation system because it is important to consider all of the items that the user might be interested. The recall does not consider the false positive rate. Since the false positive rate has not any significant value for the recommendation system, Precision and Classification accuracy have not been considered in this experiment. The evaluation of the hit ratio employs a leave-one-out methodology in which each user who interacts with the final item becomes the testing data. For example, in a test dataset of 100 items, there will be only one item that the user has interacted with. Hit ratio (HR) is a recall-based metric that does not take hit position into account. The equations of hit ratio and recall are shown in formulas 1 and 2, respectively.

$$Recall = \frac{True\ Positive}{True\ Positive + False\ Negative} \quad (1)$$

$$Hit\ Ratio = \frac{Number\ of\ hits}{Number\ of\ users} \quad (2)$$

From the previous works, it has been proven that NCF and Bert-based NCF can give the highest performance among other types of recommendation systems such as content-based recommendation systems and collaborative filtering. Thus, NCF and Bert-based NCF has been used as a baseline model to compare with the proposed BERT and CNN-based HNCF model. The experimental result and the comparison with the baseline models are shown in table 3.

TABLE III. EXPERIMENTAL RESULT ON MOVIELENS AND COMPARISON WITH THE BASELINE MODELS

| Model Name | Recall | Hit Ratio @ 10 (on 799 users) |
|---|---|---|
| Benchmark NCF [4] | 0.449 | 0.161 |
| BERT based NCF [3,7,8,9] | 0.689 | 0.422 |
| Proposed BERT-CNN based HNCF | **0.720** | **0.486** |

Table 1 shows that, shows that BERT-CNN-based HNCF model obtained Hit Ratio @ 10 of 0.486 and a recall of 0.72 for the test dataset. The proposed model outperformed the NCF and BERT-based NCF models. The sample dataset contains very few records of the original dataset. Since the previous works have shown that NCF and BERT-based NCF models can achieve high performance and the proposed BERT-CNN-based HNCF model outperformed those two models, thus it can be concluded that the proposed model can give precise and accurate recommendations.

## V. Discussion

The proposed model considers a wide range of data from the user and item. This model considers almost every piece of information that an item can have such as item features and images. It also takes consideration of the user-item interactions and extracts the latent feature from the interactions. The proposed model outperformed the state-of-the-art NCF, and BERT-based NCF models because it is considering almost every piece of information that can be generated from the users and items.

Since the images of an item pass through the pre-trained CNN model such as VGG16, the model can extract the feature vectors from the images. The categorical features of the items pass through the BERT encoder layers; thus, the model can extract the contextual meaning from those features and find out the user interests. Since the user ids and item ids are passed through the embedding layers, the model can extract the latent feature and generalize the user-item interactions. The neural network is used for classifying the user interaction with any items. By doing those steps the model can learn the user interest and recommend specific items to specific users based on their unique preferences.

## VI. Conclusion

Neural network-based collaborative filtering integrated with BERT and pre-trained CNN architecture has been introduced in this experiment. Pre-trained CNN architecture has been used instead of the general CNN model because the Pre-trained CNN model has less trainable parameters and the pre-trained parameters are trained on several multiclass images. BERT model has been used because the input sequence is converted to dense vector embeddings via the self-attention mechanism and BERT can extract contextual information from the inputs. This proposed model outperformed the baseline NCF and BERT-integrated NCF models. Thus, it can be concluded that if we consider both categorical and image data, then the performance of a recommendation system is increased efficiently.

In the future, the hyperparameters of the model can be fine-tuned using grid search or genetic algorithm to improve the model performance. The proposed takes a significant amount of time and resources to train on a large dataset. Thus, the model can optimize in terms of resource usage and time.